\documentclass[draft]{article}
\usepackage{amsmath,amsfonts,latexsym, amssymb}

\usepackage{color}

\newcommand{\const}{\mbox{const}}

\newcommand{\e}{\varepsilon}

\newcommand{\rd}{{\rm d}}
\newcommand{\bR}{{\mathbb R}}
\newcommand{\bbZ}{{\mathbb Z}}

\newcommand{\ba}{{\bf{a}}}
\newcommand{\bb}{{\bf{b}}}
\newcommand{\bx}{{\bf{x}}}
\newcommand{\by}{{\bf{y}}}
\newcommand{\bu}{{\bf{u}}}
\newcommand{\bv}{{\bf{v}}}
\newcommand{\bw}{{\bf{w}}}
\newcommand{\bz}{{\bf{z}}}

\newcommand{\bbe}{{\bf{e}}}

\newcommand{\al}{\alpha}

\newcommand{\be}{\begin{equation}}
\newcommand{\ee}{\end{equation}}

\newcommand{\cN}{{\cal N}}

\newcommand{\im}{{\text{Im} }}
\newcommand{\re}{{\text{Re} }}

\newcommand{\E}{{\mathbb E }}

\renewcommand{\P}{{\mathbb P}}
\newcommand{\bC}{{\mathbb C}}

\newcommand{\wt}{\widetilde}

\newtheorem{theorem}{Theorem}

\newtheorem{lemma}[theorem]{Lemma}

\newcommand{\qed}{\hfill\fbox{}\par\vspace{0.3mm}}
\newenvironment{proof}{{\bf Proof.}} {\hfill\qed}

\numberwithin{equation}{section}
\numberwithin{theorem}{section}
\numberwithin{definition}{section}

\title{Local semicircle law and complete
delocalization for \\ Wigner random matrices}

\author{L\'aszl\'o Erd\H os${}^1$, Benjamin
Schlein${}^1$\thanks{Supported by Sofja-Kovalevskaya Award of
the Humboldt Foundation. On leave from Cambridge University, UK}\;
and Horng-Tzer Yau${}^2$\thanks{Partially supported
by NSF grant DMS-0602038} \\
\\
Institute of Mathematics, University of Munich, \\
Theresienstr. 39, D-80333 Munich, Germany${}^1$ \\ \\
Department of Mathematics, Harvard University\\
Cambridge MA 02138, USA${}^2$ \\ \\
\\}
\begin{document}

\date{Jun 16, 2008}

\maketitle

\begin{abstract}

We consider  $N\times N$ Hermitian random matrices with  independent identical distributed
entries.  The matrix is normalized so that
the average spacing between consecutive eigenvalues is of order $1/N$.
Under suitable assumptions on the distribution
of the single matrix element, we prove that,  away from the spectral
edges,
 the density of eigenvalues concentrates around the Wigner semicircle law
on energy scales $\eta \gg N^{-1} (\log N)^8$. Up to the logarithmic
factor, this is the smallest energy scale
for which the semicircle law may be valid.
We  also prove  that for all eigenvalues away
from the spectral edges, the $\ell^\infty$-norm  of the
corresponding  eigenvectors
is of order $O(N^{-1/2})$, modulo logarithmic corrections.
The upper bound  $O(N^{-1/2})$
implies that  every eigenvector is completely delocalized, i.e.,
 the maximum size of the components of the eigenvector is of the
same order as their average size.

In the Appendix, we include a lemma by J. Bourgain which removes  one of
 our  assumptions on the distribution  of the matrix elements.
\end{abstract}


\medskip

{\it Running title:} Local semicircle law

\medskip




\section{Introduction}

The  Wigner semicircle law states that the empirical
density of the eigenvalues of a random matrix
is given by the universal semicircle distribution.
 This statement has been proved for many different ensembles,
in particular for the case when the distributions of
the entries of the matrix
are independent, identically distributed (i.i.d.).
To fix the scaling, we normalize the matrix so that
the bulk of the spectrum lies in the energy interval $[-2,2]$, i.e.,
the average spacing between consecutive eigenvalues is of order $1/N$.
We now consider a window of size $\eta$ in the bulk so that the typical
number
of eigenvalues is of order $N \eta$.
In the  usual statement of the semicircle law,
$\eta$ is a  fixed number independent of $N$
and it is taken to zero only
after the limit $N\to \infty$.
 This can be viewed as the largest scale
on which  the semicircle law is valid. On the other extreme,
for the smallest scale, one may take $\eta = k /N $ and take the limit
 $N\to \infty$ followed by $k \to \infty$.
 If the semicircle law is valid in this sense,
 we shall say that the {\it local semicircle law} holds.
 Below this smallest scale, the eigenvalue  distribution is
 expected to be governed by the Dyson statistics related to sine
kernels.
 The Dyson statistics was proved for many ensembles (see  \cite{AGZ, D}
for a review), including Wigner matrices with Gaussian convoluted
 distributions \cite{J}.

In this paper, we establish
the local semicircle law up to logarithmic factors in the energy scale, i.e.,
for
$\eta \sim  N^{-1} (\log N)^8$. The result holds for any energy
window in the bulk spectrum away from the spectral edges.
In \cite{ESY} we have proved the same statement for
$\eta \gg N^{-2/3}$ (modulo
logarithmic corrections). Prior to our work the best result was obtained
in \cite{BMT}
 for  $\eta\gg N^{-1/2}$. See also \cite{GZ} and \cite{Kh} for
related and earlier results.
As a corollary, our result also proves that no gap between consecutive
bulk
eigenvalues can be bigger than $C (\log N)^8/N$, to
be compared with the expected average $1/N$ behavior
given by Dyson's law.

It is widely believed that the eigenvalue distribution of the
Wigner random matrix
and the random Schr\"odinger operator in the
 extended (or delocalized) state regime are the same
up to normalizations. Although this conjecture
is far from the reach of the current method,
a  natural question arises as to
 whether the eigenvectors of random matrices are extended.
More precisely, if $\bv= (v_1, \ldots, v_N)$
is an $\ell^{2}$-normalized  eigenvector, $\| \bv \|=1$, we say that
$\bv$ is {\it completely delocalized}
 if $\| \bv\|_\infty = \max_j |v_j|$ is bounded from above by
$CN^{-1/2}$,  the average size of  $|v_j|$.
In this paper, we shall prove that
all eigenvectors with eigenvalues away from the spectral
edges  are completely delocalized
 (modulo logarithmic corrections) in probability.   Similar results, but
with
$C N^{-1/2}$ replaced by  $C N^{-1/3}$ were proved in  \cite{ESY}.
 Notice that our new result,
in particular, answers (up to logarithmic factors) the question
posed by T. Spencer whether $\| \bv\|_4$ is of order  $N^{-1/4}$.

\bigskip

  Denote the $(i,j)$-th entry of an $N\times N$  matrix $H$  by $h_{i,j}=h_{ij}$.
When there is no confusion, we omit the comma between the two subscripts.
We shall assume that the matrix is  Hermitian,
i.e., $h_{ij} = \overline {h_{ji}}$.
These matrices form a {\it Hermitian Wigner ensemble} if
\be
h_{i j} =   N^{-1/2} [ x_{ij} + \sqrt{-1}\;  y_{ij}],
\quad (i < j), \quad \text{and} \quad   h_{i i} =  N^{-1/2}  x_{ii},
\label{wig}
\ee
where $x_{ij},   y_{ij}$ ($i<j$) and $x_{ii}$
are independent real  random variables with  mean zero.
We assume that $x_{ij},   y_{ij}$ ($i< j$)
all have a common distribution $\nu$ with variance 1/2
and with a strictly positive density function:
$\rd \nu(x)=(\const.) e^{-g(x)}\rd x$.
The diagonal elements, $x_{ii}$,  also have  a
common distribution, $\rd\wt \nu(x)=(\const.) e^{-\wt g(x)}\rd x$,
that may be different from $\rd \nu$.
Let  $\P$ and $\E$
denote the probability and the expectation value, respectively,
 w.r.t the joint distribution of all matrix elements.

\bigskip

We need to assume further conditions
on the distributions
of the matrix elements in addition to \eqref{wig}.
\begin{itemize}
\item[{\bf C1})] The function $g$ is twice differentiable
and it satisfies
\be
  \sup_{x\in \bR} g''(x) <\infty\; .
\label{gM}
\ee
\item[{\bf C2)}] There exists a $\delta>0$ such that
\be
  \int e^{\delta x^2}\rd \wt\nu(x)  <\infty \; .
\label{x2}
\ee
\item[{\bf C4)}] The measure $\nu$ satisfies the logarithmic
Sobolev inequality, i.e., there exists a constant $C_{\text{sob}}$ such that for
any density function $u>0$ with $\int u\, \rd\nu =1$,
\be
  \int u\log u\;  \rd \nu \leq C_{\text{sob}} \int |\nabla \sqrt{u}|^2 \rd \nu\, .
\label{logsob}
\ee
\end{itemize}
Here we have followed the convention in \cite{ESY} to use the label C4)
for the logarithmic Sobolev bound and reserved  C3) for a spectral gap condition in \cite{ESY}.
We will also need the decay condition \eqref{x2}
for the measure $\rd\nu$, i.e., for some small $\delta>0$,
\be
  \int e^{\delta x^2}\rd \nu(x)  <\infty\; .
\label{x21}
\ee
 This condition was assumed in the earlier version of the manuscript, but
J.-D. Deuschel and M. Ledoux kindly pointed out to us that
\eqref{x21} follows from C4), see \cite{Le}.

Condition C1) is needed only because we will
 use Lemma 2.3 of \cite{ESY} in the  proof of the following Theorem~\ref{thm:sc}.
J. Bourgain  has informed us that this lemma
can also be proved without this condition.
We include the precise statement and his proof
in the Appendix.

\bigskip

{\it Notation.} We will use the notation $|A|$ both for
the Lebesgue measure of a set $A\subset \bR$ and  for
the cardinality of a discrete set $A\subset \bbZ$.
The usual Hermitian scalar product for vectors $\bx,\by\in \bC^N $ will be
denoted by $\bx\cdot \by$ or by $( \bx, \by)$.
We will use the convention that $C$ denotes generic large
constants and $c$ denotes generic small positive constants
whose values may change from line to line. Since we
are interested in large matrices, we always assume that $N$
is sufficiently large.

\bigskip

Let $H$ be the $N\times N$ Wigner matrix with eigenvalues
$\mu_1\leq \mu_2 \leq\ldots \leq \mu_N$.  For any spectral
parameter $z= E+i\eta\in \bC$, $\eta>0$, we denote the Green function
by $G_z= (H-z)^{-1}$. Let $F(x)=F_N(x)$ be
the empirical distribution function of the eigenvalues
\be
  F(x)= \frac{1}{N}\big| \, \big\{ \al \; : \; \mu_\al \leq x\big\}\Big|\;.
\label{Fdef}
\ee
We define the Stieltjes transform of $F$ as
\be
   m= m(z) =\frac{1}{N}\text{Tr} \; G_z = \int_\bR \frac{\rd F(x)}{x-z}\,
,
\label{Sti}
\ee
and we let
\be
\rho=\rho_{\eta}(E) = \frac{ \text{Im} \;  m(z)}{\pi}=
 \frac{1}{N\pi} \text{Im} \; \text{Tr} \; G_z
 =\frac{1}{N\pi}\sum_{\al=1}^N \frac{\eta}{(\mu_\al-E)^2+\eta^2}
\label{rhodef}
\ee
 be the normalized density of states of $H$
around energy $E$ and regularized on scale $\eta$.
The random variables $m$ and $\varrho$  also depend
on $N$, when necessary, we will indicate this fact  by
writing $m_N$ and $\varrho_N$.

For any $z=E+i\eta$ we let
$$
 m_{sc}= m_{sc}(z) = \int_\bR \frac{\varrho_{sc}(x)\rd x}{x - z}
$$
be the Stieltjes transform of the Wigner semicircle
distribution function whose density is given by
$$
  \varrho_{sc}(x) = \frac{1}{2\pi} \sqrt{4-x^2}
   {\bf 1}(|x|\leq 2)\; .
$$
For
 $\kappa, \wt\eta>0$ we define the set
$$
S_{N,\kappa,\wt\eta}:= \Big\{ z=E+i\eta\in \bC\; : \; |E|\leq 2-\kappa, \;
\wt\eta\leq \eta \leq 1\Big\}
$$
and for $\wt\eta = N^{-1}(\log N)^8$ we write
$$
S_{N,\kappa}:= \Big\{ z=E+i\eta\in \bC\; : \; |E|\leq 2-\kappa, \;
\frac{(\log N)^8}{N}\leq \eta \leq 1\Big\}.
$$
The following two theorems are the main results of this paper.

\begin{theorem}\label{thm:sc}
Let $H$ be an $N\times N$ Wigner matrix as described in \eqref{wig}
and assume the conditions \eqref{gM}, \eqref{x2} and \eqref{logsob}.
Then for any $\kappa>0$ and $\e>0$,
  the Stieltjes transform $m_N(z)$ (see \eqref{Sti})
of the empirical
eigenvalue distribution of the $N\times N$ Wigner matrix  satisfies
\be
 \P \Big\{ \sup_{z\in S_{N,\kappa}} |m_N(z)- m_{sc}(z)| \ge \e\Big\}
 \leq e^{-c(\log N)^2}
\label{mcont}
\ee
where $c>0$ depends on $\kappa, \e$. In particular, the density of states
$\varrho_\eta(E)$  converges to the
Wigner semicircle law in probability uniformly for all energies away from
the spectral edges and for all energy windows at least $N^{-1}(\log N)^8$.

Furthermore, let $\eta^*=\eta^*(N)$ such that
$(\log N)^8/N\ll\eta^*\ll 1$ as $N\to \infty$, then we
have the convergence of the counting function as well:
\be
  \P \Big\{ \sup_{|E|\leq 2-\kappa} \Big| \frac{\cN_{\eta^*}(E)}{2N\eta^*}
  - \varrho_{sc}(E)\Big|\ge \e\Big\}\leq
 e^{-c(\log N)^2}
\label{ncont}
\ee
for  any $\e>0$, where $\cN_{\eta^*}(E)= |\{ \al\; : \; |\mu_\al - E|
\leq \eta^*\}|$ denotes the number of eigenvalues in
the interval $[E-\eta^*, E+\eta^*]$.
\end{theorem}

This result identifies the density of states away from the spectral edges in a window where
the typical number of eigenvalues is of order bigger than $(\log N)^8$. Our scale is
not sufficiently small to identify individual eigenvalues, in particular we do not know
whether the local eigenvalue spacing follows the expected
 Dyson statistics characterized by the sine-kernel or some other local statistics,
e.g., that of a Poisson point process.

\begin{theorem}\label{cor:linfty} Let $H$ be an $N\times N$
Wigner matrix as described in \eqref{wig} and satisfying the
conditions (\ref{gM}), (\ref{x2})  and (\ref{logsob}).
Fix $\kappa>0$, and assume that $C$ is large enough. Then there
exists $c>0$ such that
\[
\P \Bigg\{\exists \text{ $\bv$ with $H\bv=\mu\bv$,
$\| \bv \|=1$, $\mu \in [-2+\kappa, 2-\kappa]$ and } \| \bv
\|_\infty \ge \frac{C(\log N)^{9/2}}{N^{1/2}} \Bigg\} \leq e^{-c(\log
N)^2}\;. \]
\end{theorem}

\bigskip

We now sketch the  key idea to prove Theorem \ref{thm:sc};  Theorem \ref{cor:linfty} can be proved
following similar ideas used in \cite{ESY}.

Let $B^{(k)}$ denote the $(N-1)\times(N-1)$ minor of $H$ after
removing the $k$-th row and $k$-th column and let  $m_{k}(z)$ denote the Stieltjes transform of the eigenvalue
distribution function associated with $B^{(k)}$.
It is known that  $m(z)$, defined in  \eqref{Sti},  satisfies  a recurrence relation
\be\label{recur1}
   m(z) =  \frac{1}{N}\sum_{k=1}^N \frac{1}{ h_{kk} -z -
   \big(1-\frac{1}{N}\big)m^{(k)}(z) - X_k} \; ,
\ee
where $X_{k}$ (defined precisely in \eqref{def:X}) is an ``error'' term
depending on $B^{(k)}$ and the $k$-th  column and  row elements  of the random  matrix $H$.
If we neglect $X_{k}$ (and $h_{kk}$ which is of order $N^{-1/2}$ by definition) and replace $m^{(k)}$ by $m$, we obtain an equation for
$m$ and this leads to the Stieltjes transform of the semi-circle law. So our main task is to prove that $X_{k}$ is negligible.
Unfortunately,  $X_{k}$  depends crucially on the eigenvalues and eigenfunctions of
$B^{(k)}$. In an earlier work \cite{BMT}, the estimate on $X_{k}$  was done via an involved bootstrap argument
(and valid up to order $N^{-1/2}$).   The bootstrapping is needed in \cite{BMT} since $X_{k}$ depends critically on properties of $B^{(k)}$ for which there was only limited a priori information.
In our preceding paper \cite{ESY}, we split $m$ and $m^{(k)}$ into their means and variances; the variances were then shown to be negligible up to the scale $N^{-2/3}$ (The variance control of $m$ up to the scale $N^{-1/2}$ was  already in   \cite{GZ}).  On the other hand,  the means of $m$ and
$m^{(k)}$ are very close due to the fact that the eigenvalues of $H$ and $B^{(k)}$ are  interlaced.
Finally, $X_{k}$ was controlled via an estimate on its fourth moment.
We have thus arrived at a fixed point equation  for the mean of $m$ whose unique
solution is  the
Stieltjes transform of the semi-circle law.

In the current paper, we avoid the variance control by viewing $m$ and $m^{(k)}$
directly as
random variables in the recurrence relation \eqref{recur1}. Furthermore, the moment
control on $X_{k}$ is now improved to an exponential moment estimate.  Since our
 estimate on the fourth moment of $X_{k}$ was done via a spectral gap argument,
it is a folklore that moment estimates usually can be lifted to an exponential moment
 estimate provided the spectral gap estimate is replaced by a logarithmic Sobolev
 inequality.  In this paper, we use a concentration of measure inequality
to avoid  all bootstrap arguments appearing both in \cite{BMT} and \cite{Kh}.
In the previous version of this paper we obtained the concentration inequality by
using the logarithmic Sobolev inequality together with the Gibbs entropy
inequality.   We would like to thank the referee
who pointed out  to us that the concentration
inequality of Bobkov and G\"otze \cite {BG} can be used directly to shortcut our
original proof.
The applicability of the Bobkov-G\"otze inequality as well as some of
the heuristic arguments presented here, however, depend crucially on an a
priori upper bound on $|m(z)|$;
this was obtained via  a large deviation estimate on the eigenvalue
 concentration \cite{ESY}.

\section {Proof of Theorem \ref{thm:sc}}

The proof of \eqref{ncont} follows from \eqref{mcont} exactly as
in Corollary 4.2 of \cite{ESY}, so we focus on proving
\eqref{mcont}. We first  remove the  supremum in   \eqref{mcont}.

For any two
points $z, z'\in S_{N, \kappa,\eta}$ we have
$$
  |m_N(z)-m_N(z')|\leq N^{2}|z-z'|
$$
since the gradient of $m_N(z)$ is bounded by $|\text{Im}\; z|^{-2}\leq
N^{2}$ on $S_{N, \kappa}$.
We can choose a set of at most $Q= C\e^{-2}N^{4}$ points, $z_1, z_2,
\ldots,
z_Q$, in $S_{N, \kappa,\eta}$  such that for any $z\in S_{N,\kappa,\eta}$
there exists a point $z_j$ with $|z-z_j|\leq \frac{1}{4}\e N^{-2} $.
In particular, $|m_N(z)-m_N(z_j)|\le \e/4$ if $N$ is large enough
and $|m_{sc}(z)-m_{sc}(z_j)|\leq \e/4$.
Since $\text{Im} \, z_j\ge
\eta$, under the condition that $\eta\ge N^{-1}(\log N)^8$ we have
$$
     \P \Big\{ \sup_{z\in S_{N,\kappa}} |m_N(z)-  m_{sc}(z)|
 \ge \e\Big\}
   \leq \sum_{j=1}^Q
  \P \Big\{ |m_N(z_j)-   m_{sc}(z_j)| \ge \frac{\e}{2}\Big\}
$$
Therefore, in order to conclude \eqref{mcont},
it suffices to prove that
\be
 \P \Big\{ |m_N(z)- m_{sc}(z)| \ge \e\Big\}
 \leq e^{-c(\log N)^2}
\label{mcont1}
\ee
 for each fixed $z\in S_{N,\kappa}$.

Let $B^{(k)}$ denote the $(N-1)\times(N-1)$ minor of $H$ after
removing the $k$-th row and $k$-th column. Note that
$B^{(k)}$ is an $(N-1)\times(N-1)$ Hermitian Wigner matrix
with a normalization factor off by $(1-\frac{1}{N})^{1/2}$.
 Let
$\lambda_1^{(k)}\leq \lambda_2^{(k)}\leq \ldots \leq \lambda_{N-1}^{(k)}$
denote its eigenvalues and $\bu_1^{(k)},\ldots , \bu_{N-1}^{(k)}$
the corresponding normalized eigenvectors.

Let $\ba^{(k)}=(h_{k,1}, h_{k,2}, \ldots h_{k,k-1}, h_{k,k+1}, \ldots
h_{k,N})^*
\in \bC^{N-1}$, i.e. the $k$-th column after removing the diagonal
element $h_{k,k}=h_{kk}$. Computing the $(k,k)$ diagonal element of
the resolvent $G_z$, we have
\be
   G_z(k,k)= \frac{1}{h_{kk}-z-\ba^{(k)}\cdot (B^{(k)}-z)^{-1}\ba^{(k)}}
   = \Big[ h_{kk}-z-\frac{1}{N}\sum_{\alpha=1}^{N-1}\frac{\xi_\al^{(k)}}
  {\lambda_\al^{(k)}-z}\Big]^{-1}
\label{mm}
\ee
where we defined
$$
    \xi_\al^{(k)} : = \big| \sqrt{N}\ba^{(k)}\cdot \bu_\al^{(k)}\big|^2.
$$

Similarly to the definition of $m(z)$ in \eqref{Sti},
we also define the Stieltjes transform
of the density of states of $B^{(k)}$
$$
 m^{(k)}= m^{(k)}(z) = \frac{1}{N-1}\, \text{Tr}\, \frac{1}{B^{(k)}-z}
   =\int_\bR \frac{\rd F^{(k)}(x)}{x - z}
$$
with the empirical counting function
$$
    F^{(k)}(x) = \frac{1}{N-1} \big| \, \big\{ \al \; : \;
   \lambda_{\al}^{(k)}\leq x \big\}\big|.
$$
The spectral parameter $z$ is fixed throughout the proof
and we will often omit it from the argument of the Stieltjes transforms.

It follows from \eqref{mm} that
\be
m=m(z) = \frac{1}{N}\sum_{k=1}^N G_z(k,k)
=\frac{1}{N}\sum_{k=1}^N \frac{1}{ h_{kk} - z -
\ba^{(k)} \cdot(B^{(k)}-z)^{-1} \ba^{(k)}}\, .
\label{mm1}
\ee
Let $\E_k$ denote the expectation value w.r.t
the random vector $\ba^{(k)}$.
Define the random variable
\be
  X_k(z)=X_k: =\ba^{(k)}\cdot \frac{1}{B^{(k)}-z} \ba^{(k)}
 - \E_k\; \ba^{(k)} \cdot \frac{1}{B^{(k)}-z} \ba^{(k)}
  = \frac{1}{N}\sum_{\al=1}^{N-1} \frac{\xi_\al^{(k)} -1}
   {\lambda_\al^{(k)}-z}
\label{def:X}
\ee
where we used that $\E_k \xi_\al^{(k)}=\| \bu_\al^{(k)}\|^2=1$.

We note that
$$
 \E_k\; \ba^{(k)} \cdot \frac{1}{B^{(k)}-z} \ba^{(k)}
  = \frac{1}{N}\sum_\al \frac{1}{\lambda_\al^{(k)}-z}
   = \Big(1-\frac{1}{N}\Big) m^{(k)}
$$
With this notation it follows from \eqref{mm} that
\be\label{recur}
   m =  \frac{1}{N}\sum_{k=1}^N \frac{1}{ h_{kk} -z -
   \big(1-\frac{1}{N}\big)m^{(k)} - X_k} \; .
\ee

We use that
$$
     \Big|  m -  \Big(1-\frac{1}{N}\Big)m^{(k)}\Big|
  =\Big| \int \frac{\rd F(x)}{x-z}
  -  \Big(1-\frac{1}{N}\Big)\int \frac{\rd F^{(k)}(x)}{x-z}\Big|
  = \frac{1}{N}\Big| \int \frac{NF(x)-(N-1)F^{(k)}(x)}{(x-z)^2} \rd
x\Big|.
$$
We recall that the eigenvalues of $H$ and $B^{(k)}$ are interlaced,
\be
 \mu_1\leq \lambda_1^{(k)}\leq \mu_2 \leq
\lambda_2^{(k)} \leq \ldots \leq \lambda_{N-1}^{(k)}
\leq \mu_N,
\label{interlace}
\ee
(see e.g. Lemma 2.5 of \cite{ESY}), therefore
we have $\max_x|NF(x)-(N-1)F^{(k)}(x)|\leq 1$.  Thus
\be
   \Big|  m -  \Big(1-\frac{1}{N}\Big)m^{(k)}\Big|
\leq \frac{1}{N} \int \frac{\rd x}{|x-z|^2}
  \leq  \frac{C}{N\eta}\, .
\label{mmm}
\ee

We postpone the proof of the following lemma:
\begin{lemma}\label{lm:x}
Suppose that $\bv_\alpha$ and $\lambda_\alpha$ are eigenvectors
and eigenvalues
of an $N\times N$ random matrix with a
law satisfying the assumption of Theorem \ref{thm:sc}.
Let
$$
    X = \frac{1}{N} \sum_\al \frac{\xi_\al-1}{\lambda_\al-z}
$$
with $z=E+i\eta$, $\xi_\al = |\bb\cdot \bv_\al|^2$, where
the components of $\bb$ are i.i.d.
 random variables satisfying \eqref{logsob}.
Then there exist  sufficiently small positive constants $\e_0$ and $c$
such that
in the joint product probability space of $\bb$ and the law
 of the random matrices we have
$$
   \P[ |X|\ge \e] \leq e^{- c\e (\log N)^2}
$$
for any $\e\leq \e_0$
and $\eta \ge (\log N)^8/N$.
\end{lemma}

For a given $\e>0$  and $z=E+i\eta\in S_{N, \kappa}$, we set
$z_n= E+ i2^n \eta $ and
 we define the event
$$
   \Omega = \bigcup_{k=1}^N \bigcup_{n=0}^{[\log_2 (1/\eta)]}
 \{|X_k(z_n)|\ge \e/3\}\cup \bigcup_{k=1}^N \{ |h_{kk}|\ge
\e/3\}\,,
$$
where $[\; \cdot\; ]$ denotes the integer part.
Since $h_{kk} = N^{-1/2} b_{kk}$ with $b_{kk}$  satisfying
\eqref{x2}, we have
$$
  \P \{ |h_{kk}|\ge \e/3\} \leq Ce^{-\delta\e^2 N/9}.
$$
We now apply Lemma \ref{lm:x} for each $X_k(z_n)$
and conclude that
$$
  \P(\Omega) \leq e^{- c\e (\log N)^2}
$$
with a sufficiently small $c>0$.

On the complement $\Omega^c$ we have from \eqref{recur}
$$
    m(z_n)= \frac{1}{N}\sum_{k=1}^N \frac{1}{-m(z_n) -z_n +\delta_k}
$$
where
$$
   \delta_k =\delta_k(z_n) = h_{kk} + m(z_n) - \Big(1-\frac{1}{N}\Big)
m_k(z_n) - X_k(z_n)
$$
 are random variables satisfying
 $|\delta_k|\leq \e$ by \eqref{mmm}.
After expansion, the
last equation implies that
\be
   \Big| \, m(z_n) + \frac{1}{ m(z_n) +z_n }\Big|\leq
\frac{\e}{\big( \im\, (m(z_n) + z_n) \big)
\big( \im \, (m(z_n) + z_n) - \e \big)}\,,
\label{cont1}
\ee
if $\im \, ( m(z_n) +z_n) > \e$,
using  that $ |-m(z_n)-z_n+\delta_k|\ge \im \, (m(z_n) +z_n) - \e$.

We note that for any $z\in S_{N,\kappa}$ the equation
\be
   M+ \frac{1}{M+z} =0
\label{stab1}
\ee
has a unique solution with $\text{Im} \, M>0$, namely
$M= m_{sc}(z)$, the Stieltjes transform of the semicircle law.
Note that there exists $c(\kappa)>0$ such that
$\text{Im} \, m_{sc}(E+i\eta) \ge c(\kappa)$ for any $|E|\leq 2-\kappa$,
uniformly in $\eta$.

The equation \eqref{stab1} is stable in the following sense.
For any small $\delta$, let $M=M(z,\delta)$ be a solution to
\be
   M + \frac{1}{M+z} = \delta
\label{stab2}
\ee
with $\text{Im}\, M>0$. Explicitly, we have
$$
M = \frac {-z + \sqrt {z^2 - 4 + 2 z \delta + \delta^2 } } 2 +
\frac{\delta}{2},
$$
where we have chosen the square root so that $\im M > 0$  when $\delta=0$
and $\im z > 0$. On the compact set $z\in S_{N,\kappa}$,
  $|z^2 - 4|$ is bounded away from zero
and thus
\be
   | M-m_{sc}| \leq C_\kappa \delta \,
\label{cont3}
\ee
for some constant $C_\kappa$ depending only on $\kappa$.

Now we perform a bootstrap argument in the imaginary part of $z$
 to prove that
\be
        |\, m(z) - m_{sc}(z)|\leq C^*\e
\label{cont}\ee
uniformly in $z\in S_{N,\kappa}$  with a sufficiently large constant
$C^*$. Fix $z=E+i\eta$ with $|E|\leq 2-\kappa$ and let $z_n= E+i2^n\eta$.
For $n=[\log_2 (1/\eta)]$, we have $\text{Im} \; z_n\in [\frac{1}{2}, 1]$,
 \eqref{cont} follows from \eqref{cont1} with some small $\e$, since
the right hand side of \eqref{cont1} is bounded by $C\e$.
Suppose now that \eqref{cont} has been proven for $z=z_n$, for some $n\geq 1$ with
$\eta_n = \text{Im}\; z_n \in [2N^{-1}(\log N)^8,\, 1]$. We
want to prove it for $z=z_{n-1}$, with $\text{Im}\; z_{n-1}=\eta_n/2$.
By integrating the inequality
$$
    \frac{\eta_n/2}{(x-E)^2 + (\eta_n/2)^2}
\ge \frac{1}{2} \frac{\eta_n}{(x-E)^2+\eta_n^2}
$$
with respect to $\rd F(x)$ we obtain that
$$
 \text{Im}\, m(z_{n-1}) \ge \frac{1}{2} \text{Im}\,
  m(z_n)
\ge \frac{1}{2}c(\kappa)- C^*\e > \frac{c(\kappa)}{4}\,
$$
for sufficiently small $\e$, where \eqref{cont}
and $\im\, m_{sc}(z_n)\ge c(\kappa)$ were
used. Thus the right hand side of \eqref{cont1}
with $z_n$ replaced with $z_{n-1}$
is bounded by $C\e$, the constant depending only on $\kappa$.
Applying the stability bound \eqref{cont3}, we get \eqref{cont}
for $z=z_{n-1}$. Continuing the induction argument,
finally we obtain  \eqref{cont} for $z=z_0=E+i\eta$. \qed

\section{ Proof of Lemma \ref{lm:x}}

Let $I_n = [n\eta, (n+1)\eta]$ and $K_0$
be a sufficiently  large number.
We have $[-K_0, K_0] \subset \cup_{ n = -m}^m  I_n$ with $m \le
CK_0/\eta$.
Denote by $\Omega$ the event
$$
  \Omega : = \Big\{ \max_n \cN_{I_n} \ge N\eta (\log N)^2\Big\}
\cup \{ \max_\al |\lambda_\al|\ge K_0\}
$$
where $\cN_{I_n}=|\{\al\; : \; \lambda_\al\in I_n\}|$
 is the number of eigenvalues in the interval $I_n$.
{F}rom Theorem 2.1 and Lemma 7.4  of \cite{ESY},
  the probability of $\Omega$ is bounded by
$$
   \P(\Omega) \leq e^{-c(\log N)^2}.
$$
 for some sufficiently small $c>0$.
Therefore, if $\P_\bb$ denotes the probability w.r.t. the variable $\bb$, we find
\be
\begin{split}
  \P[ |X|\ge \e] & \leq e^{-c(\log N)^2} + \E \Big[ {\bf 1}_{\Omega^c}
   \P_\bb  [ |X|\ge \e] \Big]\\
   & \leq  e^{-c(\log N)^2} + \E \Big[ {\bf 1}_{\Omega^c} \,
   \cdot \P_\bb [ \text{Re} X \le - \e /2  ] \Big] +
 \E \Big[ {\bf 1}_{\Omega^c} \, \cdot
 \P_\bb [ \text{Re} X \ge  \e /2  ] \Big] \\ &
 \hspace{.5cm} +  \E \Big[ {\bf 1}_{\Omega^c} \,
 \cdot \P_\bb [ \text{Im} X \le - \e /2  ] \Big] +
  \E \Big[ {\bf 1}_{\Omega^c} \,   \cdot \P_\bb [ \text{Im} X \ge \e /2  ] \Big].
\end{split}\label{T}
\ee
The last four terms on the r.h.s. of the last equation
can all be handled with similar arguments; we show, for example,
how to bound the last term. For any $T>0$, we have
\begin{equation}\label{eq:T2} \E \Big[ {\bf 1}_{\Omega^c} \,
 \cdot \P_\bb [ \text{Im} X \ge \e /2  ] \Big] \leq e^{- T \e / 2} \,
 \E \Big[ {\bf 1}_{\Omega^c} \, \E_\bb \, e^{T \, \text{Im} X} \Big] \end{equation}
where $\E_\bb$ denotes the expectation w.r.t. the variable $\bb$.
Using the fact that the distribution of the components of $\bb$
satisfies the log-Sobolev inequality (\ref{logsob}), it follows
from the concentration inequality (Theorem 2.1 from \cite{BG}) that
\begin{equation} \E_{\bb} \, e^{T \, \text{Im} X} \leq  \E_{\bb} \,
\exp \left( \frac{C_{\text{sob}} T^2}{2} \, |\nabla ( \text{Im} X)|^2 \right) \,
 \end{equation}
where
\begin{equation}
\begin{split}
|\nabla ( \text{Im} X)|^2 = \; & \sum_k  \Big( \Big|
 \frac{\partial \, (\text{Im} X)}{\partial\, (\re \, b_k)}\Big|^2 +
\Big| \frac{\partial \, (\text{Im} X)}{\partial \, ( \text{Im} \, b_k)} \Big|^2 \Big) \\
=\; & \sum_k \Big( \Big| \frac{\eta}{N} \sum_{\al} \frac{1}{|\lambda_{\al} - z|^2}
 \left( (\bb \cdot \bv_{\al}) \, \overline{\bv}_{\al} (k) +
(\overline{\bb \cdot \bv_{\al}}) \, \bv_{\al} (k) \right) \Big|^2 \\ &
\hspace{2cm}+ \Big| \frac{\eta}{N} \sum_{\al} \frac{1}{|\lambda_{\al} - z|^2}
 \left( (\bb \cdot \bv_{\al})  \overline{\bv}_{\al} (k) - (\overline{\bb \cdot \bv_{\al}}) \,
 \bv_{\al} (k) \right) \Big|^2 \Big) \\
= \; &4 \frac{\eta^2}{N^2} \sum_{\al} \frac{\xi_{\alpha}}{|\lambda_{\alpha} - z|^4} \\
\leq \; & \frac{4}{N\eta} \, Y \,.
\end{split}
\end{equation}
Here we defined the random variable
\[ Y = \frac{1}{N} \sum_{\al}  \frac{\xi_\al}{|\lambda_\al -z|} \, .\]
{F}rom (\ref{eq:T2}), choosing $T = 2 (\log N)^2$ and using that
 $N\eta \geq (\log N)^8$, we obtain
\begin{equation}\label{eq:Y2}
\E \Big[ {\bf 1}_{\Omega^c} \,   \cdot \P_\bb [ \text{Im} X \ge \e /2  ]
 \Big] \leq e^{- \e (\log N)^2} \, \E \Big[ {\bf 1}_{\Omega^c} \,
\E_{\bb} \, \exp \left( \frac{8C_{\text{sob}}}{(\log N)^4} \, Y \right) \Big]\,.
\end{equation}
Let $\nu = 8C_{\text{sob}}/(\log N)^4$.
By H\"older inequality, we can estimate
\be
  \E_\bb e^{\nu Y} = \E_\bb \prod_\al
\exp{\Big[\frac{\nu }{N|\lambda_\al -z|}\xi_\al\Big]}
\leq      \prod_\al  \Bigg( \E_\bb
\exp{\Big[\frac{\nu c_\al}{N|\lambda_\al -z|}\xi_\al\Big]}
\Bigg)^{1/c_\al},
\label{hold}
\ee
where $\sum_\al \frac{1}{c_\al}=1$. We shall choose
$$
   c_\al = \varrho \frac{N|\lambda_\al -z|}{\nu}
$$
where $\varrho$ is given by
$$
      \varrho = \frac{\nu}{N}\sum_\al \frac{1}{|\lambda_\al -z|} \leq
      \frac{\nu \log N}{N\eta}\max_n \cN_{I_n} \leq \nu (\log N)^3
    = \frac{8C_{\text{sob}}}{\log N} \, .
$$
Here we have used  $\max_n \cN_{I_n} \le N \eta (\log N)^2$
due to that we are in the set  $\Omega^c$.
Notice that with this choice,
$$
\frac{ \nu c_\al}{N |\lambda_\al -z|}\leq\frac{8C_{\text{sob}}}{\log N}
$$
is a small number. In the proof of Lemma 7.4 of \cite{ESY}
(see equation (7.13) of \cite{ESY})  we
showed that
$$
     \E_\bb \; e^{\tau \xi_\al} < K
$$
with a universal constant $K$
if $\tau$ is sufficiently small depending on $\delta$ in \eqref{x2}.
{F}rom (\ref{eq:Y2}), it follows that
\begin{equation}
\E \Big[ {\bf 1}_{\Omega^c} \,   \cdot \P_\bb [ \text{Im} X \ge \e /2  ] \Big]
 \leq K e^{- \e (\log N)^2}
\end{equation}
Since similar bounds hold for the other terms on the r.h.s. of (\ref{T}) as well,
this concludes the proof of the lemma.
\qed

\section{Delocalization of eigenvectors}

\bigskip

Here we prove Theorem \ref{cor:linfty},
the argument follows the same line as in \cite{ESY}
(Proposition 5.3).
Let $\eta^* = N^{-1} (\log N)^9$ and partition the
interval $[-2+\kappa, 2-\kappa]$ into $n_0= O(1/\eta^*)\leq O(N)$
intervals $I_1, I_2, \ldots I_{n_0}$
of length $\eta^*$.
As before, let $\cN_{I}=|\{ \beta\; : \; \mu_\beta\in I\}|$
denote the eigenvalues in $I$.
 By using \eqref{ncont} in Theorem \ref{thm:sc},
we have
\[
\P \left\{\max_n \cN_{I_n} \leq \e N\eta^* \right\}
 \leq e^{-c (\log N)^2}.
\]
if $\e$ is sufficiently small (depending on $\kappa$).
Suppose that $\mu
\in I_n$, and that $H\bv = \mu \bv$. Consider the decomposition
\be
\label{Hd} H = \begin{pmatrix} h & \ba^* \\
\ba & B
\end{pmatrix}
\ee
where  $\ba= (h_{1,2}, \dots h_{1,N})^*$  and  $B$ is the $(N-1)
\times (N-1)$ matrix obtained by removing the first row and first column
from $H$. Let  $\lambda_\al$ and $\bu_\al$ (for $\al=1,2,\ldots , N-1$)
 denote the eigenvalues and the normalized eigenvectors  of $B$.
 {F}rom the eigenvalue equation $H \bv = \mu \bv$
and from \eqref{Hd}
 we find that
\be\label{ee}
h v_1 + \ba \cdot \bw = \mu v_1, \quad \text{and } \quad \ba v_1 + B
\bw = \mu \bw
\ee
with $\bw= (v_2, \dots ,v_N)^t$. {F}rom these equations
we obtain
$ \bw = (\mu-B)^{-1} \ba v_1 $ and thus
$$
\|\bw\|^2= \bw\cdot \bw = |v_1|^2 \ba\cdot (\mu -B)^{-2} \ba
$$
Since $\|\bw\|^2 = 1 - |v_1|^2$, we obtain
\be\label{v2} |v_1|^2 =
\frac{1}{1+ \ba\cdot(\mu -B)^{-2} \ba} = \frac{1}{1 + \frac{1}{N}
\sum_{\alpha} \frac{\xi_{\alpha}}{(\mu - \lambda_{\alpha})^2}}
\leq \frac{4 N [\eta^*]^2}{\sum_{\lambda_\alpha \in I_n} \xi_{\alpha}} \,
,
\ee
where in the second equality we set
$\xi_{\alpha} = |\sqrt{N} \ba \cdot \bu_{\alpha}|^2$
 and used
the spectral representation of $B$. By the interlacing property
of  the eigenvalues of $H$ and $B$,
there exist at least $\cN_{I_n}-1$ eigenvalues $\lambda_\al$
in $I_n$.
Therefore, using that the components of any eigenvector are identically
distributed, we have
\begin{equation}
\begin{split}
\P \Big( \exists &\text{ $\bv$ with $H\bv=\mu\bv$, $\| \bv \|=1$,
$\mu \in [-2+\kappa, 2-\kappa]$ and } \| \bv \|_\infty \ge
\frac{C(\log N)^{9/2}}{N^{1/2}} \Big)  \\
&\leq N n_0 \sup_n \P \Big( \exists \text{ $\bv$ with
$H\bv=\mu\bv$, $\| \bv \|=1$, $\mu \in I_n$ and } |v_1|^2 \ge
\frac{C(\log N)^9}{N} \Big)\\
&\leq \const \,  N^2 \sup_n \P \left( \sum_{\lambda_\alpha \in I_n}
\xi_{\alpha} \leq \frac{4N\eta^*}{C} \right) \\
&\leq  \const \, N^2 \sup_n \P \left( \sum_{\lambda_\alpha \in I_n}
\xi_{\alpha} \leq \frac{4 N\eta^*}{C} \text{ and } \cN_{I_n} \geq
\e N\eta^*\right) + \const \,  N^2  \sup_n \, \P
\left(\cN_{I_n} \leq \e N\eta^* \right)
\\&\leq \const \,  N^2 e^{-c(\log N)^9} + \const \, N^2 e^{-c (\log
N)^2} \leq e^{-c' (\log N)^2}\,,
\end{split}
\end{equation}
by choosing $C$ sufficiently large,
depending on $\kappa$ via $\e$. Here we used
 Corollary 2.4. of \cite{ESY} that states that under condition C1) in \eqref{gM}
there exists a positive $c$ such that for any $\delta$ small enough
\begin{equation}\label{ld}
     \P \left( \sum_{\alpha \in A} \xi_\alpha \leq \delta
 m \right) \le e^{- c m}\;
\end{equation}
for all $A \subset \{1, \cdots, N-1 \}$ with cardinality $|A|=m$.
We remark that by applying Lemma \ref{lm:BL} from the Appendix instead of
Lemma 2.3 in \cite{ESY}, the bound \eqref{ld} also holds without
condition C1) if the matrix elements are bounded random variables.
It is clear that the boundedness assumption in Lemma  \ref{lm:BL} can be relaxed
by performing an appropriate cutoff argument; we will not pursue this direction
in this article.
\qed



\bigskip

\centerline{\Large\bf{Appendix: Removal of the assumption C1)}}


\bigskip

\centerline{\Large Jean Bourgain}
\centerline{\large School of Mathematics}
\centerline{\large Institute for Advanced Study}
\centerline{\large Princeton, NJ 08540, USA}

\bigskip

The following Lemma shows that  the assumption C1)
in Lemma 2.3 and its corollary
in \cite{ESY}  can be removed.

\begin{lemma}\label{lm:BL} Suppose that $z_1, \dots ,z_N$ are bounded,
complex valued i.i.d. random variables with $\E \, z_i = 0$ and
$\E \, |z_i|^2 = a>0$.
 Let $P: \bC^N \to \bC^N$ be a rank-$m$ projection,
and $\bz = (z_1, \dots ,z_N)$. Then, if $\delta$
is small enough, there exists $c>0$ such that
\[ \P \, \left( |P \bz|^2 \leq \delta m \right) \leq e^{-cm} \, .\]
\end{lemma}

\medskip

Lemma 2.3 in \cite{ESY} stated that the same conclusion holds under the
condition C1), but it required no assumption on the boundedness of the
random variables.

\medskip

\begin{proof}
It is enough to prove that
\begin{equation}\label{eq:clb} \P \left( \left| |P \bz|^2 - a m \right| >
 \tau m \right) \leq e^{-c \tau^2 m} \, \end{equation}
for all $\tau$ sufficiently small. Introduce the notation
$\| X \|_q=\big[\E |X|^q\big]^{1/q}$. Since
\begin{equation}
\begin{split}
\P \left( \left| |P\bz|^2 - a m \right| > \tau m \right)
\leq \frac{\left\| \, |P \bz|^2 - a m \right\|_q^q}{(\tau m)^q},
\end{split}
\end{equation}
the bound (\ref{eq:clb}) follows by showing that
\begin{equation}\label{eq:clb2} \| | P \bz |^2 - a m \|_q \leq
 C \sqrt{q} \, \sqrt{m} \qquad \text{for all $q < m$} \end{equation}
(and then choosing $q = \alpha \tau^2 m$ with a small
enough $\alpha$). To prove (\ref{eq:clb2}), observe that
 (with the notation $\bbe_i = ( 0, \dots , 0 ,1 , 0 \dots, 0)$
 for the standard basis of $\bC^N$)
\begin{equation}\label{eq:clb3}
\begin{split}
| P \bz |^2 &= \sum_{i=1}^N |z_i|^2 \, |P \bbe_i|^2 + \sum_{i \neq j}^N
\overline{z}_i \, z_j \, P \bbe_i \cdot P \bbe_j = a m + \sum_{i=1}^N
\left( |z_i|^2 - \E |z_i|^2 \right) \, |P \bbe_i|^2 + \sum_{i\neq j}
 \overline{z}_i z_j \, P \bbe_i \cdot P \bbe_j
\end{split}
\end{equation}
and thus
\begin{equation}\label{eq:clb4}
\begin{split}
\left\| |P \bz|^2 - a m \right\|_q \leq \left\| \, \sum_{i=1}^N
 \left( |z_i|^2 - \E |z_i|^2 \right) |P \bbe_i|^2 \right\|_q + \left\| \,
\sum_{i \neq j}^N \overline{z}_i z_j \, P \bbe_i \cdot P \bbe_j \right\|_q
\,.
\end{split}
\end{equation}
To bound the first term, we use that for arbitrary i.i.d. random variables
 $x_1, \dots ,x_N$ with $\E \, x_j =0$ and
$\E \, e^{\delta |x_j|^2} < \infty$ for some $\delta >0$, we have the
bound
\begin{equation}\label{eq:bern}
\| X \|_q \leq C \sqrt{q} \, \| X \|_2 \end{equation} for $X =
\sum_{j=1}^N a_j x_j$, for arbitrary $a_j \in \bC$.
The bound (\ref{eq:bern}) is an extension of Khintchine's inequality
 and it can be proven as follows using the representation
\begin{equation}\label{eq:bern1}
  \| X \|_q^q = q\int_0^\infty   \rd y  \; y^{q-1} P( |X|\ge y) \, .
\end{equation}
Writing $a_j = |a_j|e^{i\theta_j}$, $\theta_j\in \bR$, and decomposing
$e^{i\theta_j}x_j$ into real and imaginary parts,
it is clearly sufficient to prove \eqref{eq:bern} for the case when
$a_j, x_j\in \bR$ are real and $x_j$'s are independent with $\E \, x_j =0$
and
$\E \, e^{\delta |x_j|^2} < \infty$.
To bound the probability $\P (|X| \geq y)$ we observe that
\[    \P(X\ge y) \leq e^{-ty} \, \E \, e^{tX} = e^{-t y} \, \prod_{j=1}^N
  \E \, e^{t a_j x_j} \leq e^{-t y} \, e^{C t^2\sum_{j=1}^N a_j^2} \]
because $\E  \, e^{\tau x} \leq e^{C \tau^2}$  from the moment assumptions
on $x_j$
with a sufficiently large $C$ depending on
$\delta$. Repeating this argument for $- X$, we find
\[ \P(|X|\ge y) \leq 2 \, e^{-ty} \, e^{Ct^2\sum_{j=1}^N a_j^2}
 \leq e^{-y^2/(2 C \sum_{j=1}^N a_j^2)} \] after optimizing in $t$.
The estimate (\ref{eq:bern}) follows then by plugging the last bound
into (\ref{eq:bern1}) and computing the integral.

\medskip

Applying (\ref{eq:bern}) with $x_i = |z_i|^2 - \E \, |z_i|^2$
($\E \, e^{\delta x_i^2} < \infty$ follows from the assumption
$\| z_i \|_{\infty} < \infty$), the first term on the r.h.s. of
(\ref{eq:clb4}) can be controlled by
\begin{equation}\label{eq:clb5}
\left\| \, \sum_{i=1}^N \left( |z_i|^2 - \E |z_i|^2 \right)
 |P \bbe_i|^2 \right\|_q \leq C \sqrt{q} \left( \sum_{i=1}^N
 |P \bbe_i |^4 \right)^{1/2} \leq C \sqrt{q} \left(
 \sum_{i=1}^N |P \bbe_i |^2 \right)^{1/2} = C \sqrt{q} \sqrt{m}\,.
\end{equation}
As for the second term on the r.h.s. of (\ref{eq:clb4}), we define
the functions $\xi_j (s), s \in [0,1], j=1,\dots, N$ by
\[ \xi_j (s) =  \left\{  \begin{array}{ll} 1 \quad &\text{if }
 s \in \bigcup_{k=0}^{2^{j-1} - 1} \, \left[ \frac{2k}{2^j},
\frac{2k+1}{2^j} \right) \\ 0 \quad &\text{otherwise} \end{array} \right.
\, .\]
 Since
\[ \int_0^1 \rd s \; \xi_i (s) (1- \xi_j (s)) = \frac{1}{4} \] for all
 $i \neq j$, the second term on the r.h.s. of (\ref{eq:clb4}) can be
 estimated by
\begin{equation}\label{eq:clb6}
\left\| \, \sum_{i \neq j}^N \overline{z}_i z_j \, P \bbe_i \cdot
P \bbe_j \right\|_q \leq  4 \int_0^1 \rd s \; \left\| \, \sum_{i \neq j}^N
 \xi_i (s) \, (1- \xi_j (s)) \, \overline{z}_i z_j \,
 P \bbe_i \cdot P \bbe_j \right\|_q \,.
\end{equation}
For fixed $s \in [0,1]$, set \[ I (s) = \{ 1 \leq i
 \leq N : \xi_i (s) = 1 \} \qquad \text{and } \quad J (s) = \{ 1, \dots ,N
\}
\backslash I (s) \, . \] Then
\begin{equation*}
\left\|\, \sum_{i \neq j}^N  \xi_i (s) \, (1- \xi_j (s))
 \, \overline{z}_i z_j \, P
\bbe_i \cdot P \bbe_j \right\|_q = \left\| \,
\sum_{i \in I (s), j \in J (s)} \overline{z}_i z_j \, P \bbe_i
\cdot P \bbe_j  \right\|_q =
 \left\| \, \sum_{j \in J(s)} z_j \left( \sum_{i \in I(s)}
 z_i P \bbe_i \right)
\cdot \bbe_j \right\|_q \, .
\end{equation*}
Since by definition $I \cap J = \emptyset$,
the variable $\{ z_i \}_{i \in I}$ and the variable
 $\{ z_j \}_{j \in J}$ are independent. Therefore,
 we can apply Khintchine's inequality (\ref{eq:bern})
 in the variables $\{ z_j \}_{j \in J}$
 (separating the real and imaginary parts) to conclude that
\begin{equation}
\begin{split}
\left\|\, \sum_{i \neq j}^N  \xi_i (s) \, (1- \xi_j (s)) \,
 \overline{z}_i z_j \, P
\bbe_i \cdot P \bbe_j \right\|_q   &\leq C \sqrt{q}
\left\| \, \left( \, \sum_{j \in J(s)} \left|
\left(\sum_{i \in I (s)} z_i P \bbe_i \right) \cdot
 \bbe_j \right|^2 \right)^{1/2} \right\|_q \\ & \leq C \sqrt{q}
\, \left\| \, \sum_{i \in I (s)} z_i P \bbe_i \right\|_q
\leq C \sqrt{q} \, \| P \bz \|_q  \,
\end{split}
\end{equation}
for every $s \in [0,1]$. It follows from (\ref{eq:clb6}) that
\[  \left\| \, \sum_{i \neq j}^N \overline{z}_i z_j \, P \bbe_i
\cdot P \bbe_j \right\|_q \leq C \, \sqrt{q} \, \| P\bz \|_q \, . \]
 Inserting the last equation and (\ref{eq:clb5}) into the r.h.s. of
 (\ref{eq:clb4}), it follows that
\[ \left\| |P \bz|^2 - a m \right\|_q \leq C \, \sqrt{q} \,
\left( \sqrt{m} + \| P \bz \|_q \right) \, . \]
Since clearly \[  \| P \bz \|_q \leq \left\|
|P \bz|^2 - a m \right\|^{1/2}_q + \sqrt{am} \]
the bound (\ref{eq:clb2}) follows immediately.
\end{proof}

\bigskip

{\it Acknowledgments}: We thank the referee for very useful
comments on earlier versions of this paper.
We also thank J. Bourgain for the kind permission
to include his result in the appendix.


\thebibliography{hhh}


\bibitem{AGZ} Anderson, G. W.,  Guionnet, A., Zeitouni, O.:
Lecture notes on random matrices. Book in preparation.


\bibitem{BMT} Bai, Z. D., Miao, B.,
 Tsay, J.: Convergence rates of the spectral distributions
 of large Wigner matrices.  {\it Int. Math. J.}  {\bf 1}
  (2002),  no. 1, 65--90.

\bibitem{BG}
Bobkov, S. G., G\"otze, F.: Exponential integrability
and transportation cost related to logarithmic
Sobolev inequalities. {\it J. Funct. Anal.} {\bf 163} (1999), no. 1, 1--28.


\bibitem{D} Deift, P.: Orthogonal polynomials and
random matrices: a Riemann-Hilbert approach.
{\it Courant Lecture Notes in Mathematics} {\bf 3},
American Mathematical Society, Providence, RI, 1999.





\bibitem{ESY} Erd{\H o}s, L., Schlein, B., Yau, H.-T.:
Semicircle law on short scales and delocalization
of eigenvectors for Wigner random matrices.
2007, preprint. {arXiv.org:0711.1730}.


\bibitem{GZ} Guionnet, A., Zeitouni, O.:
Concentration of the spectral measure
for large matrices. {\it Electronic Comm. in Probability}
{\bf 5} (2000), Paper 14.


\bibitem{J} Johansson, K.: Universality of the local spacing
distribution in certain ensembles of Hermitian Wigner matrices.
{\it Comm. Math. Phys.} {\bf 215} (2001), no.3. 683--705.

\bibitem{Kh} Khorunzhy, A.: On smoothed density
of states for Wigner random matrices. {\it Random Oper.
Stoch. Eq.} {\bf 5} (1997), no.2., 147--162.

\bibitem{Le} Ledoux, M.: The concentration of measure phenomenon.
Mathematical Surveys and Monographs, {\bf 89} American Mathematical Society, Providence, RI, 2001.


\bibitem{QY}  Quastel, J,  Yau, H.-T.:   Lattice gases, large deviations, and the incompressible
  Navier-Stokes equations,   Ann. Math, {\bf 148}, 51-108, 1998.
\end{document}